\documentclass{article}

\usepackage{booktabs}
\usepackage{graphics}
\usepackage{graphicx}
\usepackage{listings}
\usepackage{nameref}
\usepackage{algorithm}
\usepackage{algpseudocode}
\algrenewcommand\algorithmicrequire{\textbf{Input:}}
\algrenewcommand\algorithmicensure{\textbf{Output:}}
\usepackage{bm}
\usepackage{amsmath}
\usepackage{amssymb}
\usepackage{caption} 
\usepackage{multirow}

\usepackage{arxiv}

\usepackage[utf8]{inputenc} 
\usepackage[T1]{fontenc}    
\usepackage{hyperref}       
\usepackage{url}            
\usepackage{booktabs}       
\usepackage{amsfonts}       
\usepackage{nicefrac}       
\usepackage{microtype}      
\usepackage{lipsum}
\usepackage{graphicx}
\graphicspath{ {./images/} }

\title{PhaTYP: Predicting the lifestyle for bacteriophages using BERT}

\author{
 Jiayu Shang \\
  Dept. of Electrical Engineering\\
  City University of Hong Kong\\
  Kowloon, Hong Kong SAR, China\\
  \texttt{jyshang2-c@my.cityu.edu.hk} \\
  \And
 Xubo Tang \\
  Dept. of Electrical Engineering\\
  City University of Hong Kong\\
  Kowloon, Hong Kong SAR, China\\
  \texttt{xubotang2-c@my.cityu.edu.hk} \\
  \And
 Yanni Sun \\
  Dept. of Electrical Engineering\\
  City University of Hong Kong\\
  Kowloon, Hong Kong SAR, China\\
  \texttt{yannisun@cityu.edu.hk} \\
}

\begin{document}

\maketitle
\begin{abstract}
Bacteriophages (or phages), which infect bacteria, have two distinct lifestyles: virulent and temperate. Predicting the lifestyle of phages helps decipher their interactions with their bacterial hosts, aiding phages' applications in fields such as phage therapy. Because experimental methods for annotating the lifestyle of phages cannot keep pace with the fast accumulation of sequenced phages, computational method for predicting phages' lifestyles has become an attractive alternative. Despite some promising results, computational lifestyle prediction remains difficult because of the limited known annotations and the sheer amount of sequenced phage contigs assembled from metagenomic data. In particular, most of the existing tools cannot precisely predict phages' lifestyles for short contigs. In this work, we develop PhaTYP (Phage TYPe prediction tool) to improve the accuracy of lifestyle prediction on short contigs. We design two different training tasks, self-supervised and fine-tuning tasks, to overcome lifestyle prediction difficulties. We rigorously tested and compared PhaTYP with four state-of-the-art methods: DeePhage, PHACTS, PhagePred, and BACPHLIP. The experimental results show that PhaTYP outperforms all these methods and achieves more stable performance on short contigs. In addition, we demonstrated the utility of PhaTYP for analyzing the phage lifestyle on human neonates' gut data. This application shows that PhaTYP is a useful means for studying phages in metagenomic data and helps extend our understanding of microbial communities.
\end{abstract}

\section{Introduction}
\label{sec:intro}
Bacteriophages (aka phages) are viruses that infect bacteria. They are widely regarded as the most abundant and diverse entities in the biosphere \cite{mcgrath2007bacteriophage} and play an essential role in various ecosystems \cite{zhong2021glacier, nishimura2017environmental}. For example, by lysing the bacterial host, phages can regulate both the composition and function of the microbiome. With the in-depth study of phages, there is accumulating evidence revealing phages' significant impacts on various fields, such as dairy production \cite{moineau1999applications, brussow2001comparative}, phage therapy \cite{azimi2019phage, loc2011pros}, and disease diagnostics \cite{wang2004epitope,bazan2012phage}.

Phages' applications depend on annotations of their lifestyles. There are two types of phages based on their lifestyles: virulent phages and temperate phages. 
Virulent phages infect bacteria and kill their hosts to release their offsprings. But they don't integrate their genomes into the hosts \cite{shkoporov2018varphicrass001}. In contrast, temperate phages integrate their genomes into the host chromosome and copy their genomes together with the host \cite{mirzaei2017menage}. They will maintain this living state, which is also called prophage, until induced by appropriate conditions and enter the lytic cycle to kill their hosts \cite{clarke1998virus, clark1986effects}. The lifestyle of the phages can directly affect their usages. For example, virulent phages are required in phage therapy to kill the antibiotic-resistant bacteria \cite{housby2009phage, brives2020phage}, while temperate phages can engineer a host’s genome \cite{menouni2015bacterial} and help regulate gene expression and change cell physiology by introducing novel functions \cite{feiner2015new, howard2017lysogeny}. However, culturing and isolating phages in lab for identifying the lifestyles are usually expensive and time-consuming \cite{mcnair2012phacts, camarillo2021massive}, especially for phages infecting anaerobes, such as \textit{Clostridioides difficile} and \textit{Mycobacterium tuberculosis} \cite{hargreaves2014clostridium, xiong2014titer, carrigy2019prophylaxis}. Metagenomics allows sequencing of uncultured dark matter of the microbial biosphere, which can contain a large number of phages \cite{marine16}. Being able to annotate the lifestyles of phages sequenced from host-associated or natural environments is expected to extend our knowledge about phage composition and their interactions with other microbes. Thus, computational prediction of the phage lifestyles has become an attractive alternative to experimental methods.

There are two main challenges for computational prediction of the lifestyle of phages. First, the number of reference phages with known lifestyle annotations is very limited. According to the latest lifestyle annotation dataset provided by \cite{wu2021deephage}, there are 1,290 virulent phages and 577 temperate phages. However, the number of released phages in the RefSeq database is 4,517 in 2021, indicating that over a half of phages have no annotations. An even larger data source for phage is IMG/VR v3 database \cite{IMGVR}, which contains nearly 2 million uncultivated phage-like genomes in 2021. Because of the limited number of annotated genomes, most phages cannot be classified by sequence match-based methods. Second, as mobile genetic elements, phages usually mobilize host genetic material and incorporate it into their own genomes \cite{edwards2016computational}, leading to poor sequence assembly results and incomplete fragments for phages \cite{wu2021deephage}. Both the short length of the fragments and the ambiguous regions increase the difficulty of the lifestyle prediction task.

\subsection{Related work}
\label{sec:relate}
Given the importance of phages, numerous efforts have been made to computationally predict phages' taxonomic labels \cite{pons2021vpf, bin2019taxonomic, shang2021bacteriophage} and their hosts \cite{shang2021predicting, amgarten2020vhulk, Nathan2020vhmnet, shang2022cherry}, and to identify phages from metagenomic data \cite{shang2022accurate, VirFinder, VirFinder}. In addition, there are a handful of tools for phage lifestyle prediction \cite{mcnair2012phacts, hockenberry2021bacphlip, wu2021deephage}. One type of method uses maker genes to distinguish virulent and temperate phages. For example, integrase and excisionase are two widely accepted marker genes for identifying temperate phages \cite{emerson2012dynamic}. However, only a few genes can be used as marker genes, especially for virulent phages \cite{mcnair2012phacts}. In addition, the fragmented contigs from the metagenomic assembly may not cover such genes, and thus, using a small set of marker genes can lead to a low recall for lifestyle classification.

Instead of relying on handcrafted marker genes, learning-based methods are proposed aiming to automatically learn features from two types of phages' DNA and protein sequences. For example, PHACTS \cite{mcnair2012phacts} trained a random forest model on protein similarities for virulent and temperate phage classification. BACPHLIP \cite{hockenberry2021bacphlip} trained another random forest classifier using a set of lysogeny-associated protein domains identified by HMMER3 \cite{eddy2011accelerated}. However, such strategies may not apply to metagenomic data. According to the benchmark results shown in \cite{wu2021deephage}, the accuracy of PHACTS decreases on short contigs. For example, PHACTS only achieves an accuracy of \textasciitilde60\% when the phage contigs are below 2kbp. Also, BACPHLIP is only designed for complete phage genomes according to the provided guidelines; it returns errors when short contigs are provided as inputs. Unlike these methods, PhagePred \cite{song2020classifying} and DeePhage \cite{wu2021deephage} can identify the lifestyle for contigs assembled from metagenomic data. PhagePred calculated the distance between a query and a reference using a learned $k$-mer frequency-based Markov model. DeePhage \cite{wu2021deephage}, which has the best reported performance on lifestyle prediction, can predict contigs as short as 100bp. DeePhage distinguishes the lifestyle of phages by applying a convolutional neural network to learn the motif-related information from DNA sequences. Nevertheless, its best performance on the short contigs is only \textasciitilde80\%.

\begin{figure*}[h!]
    \centering
    \includegraphics[width=0.8\linewidth]{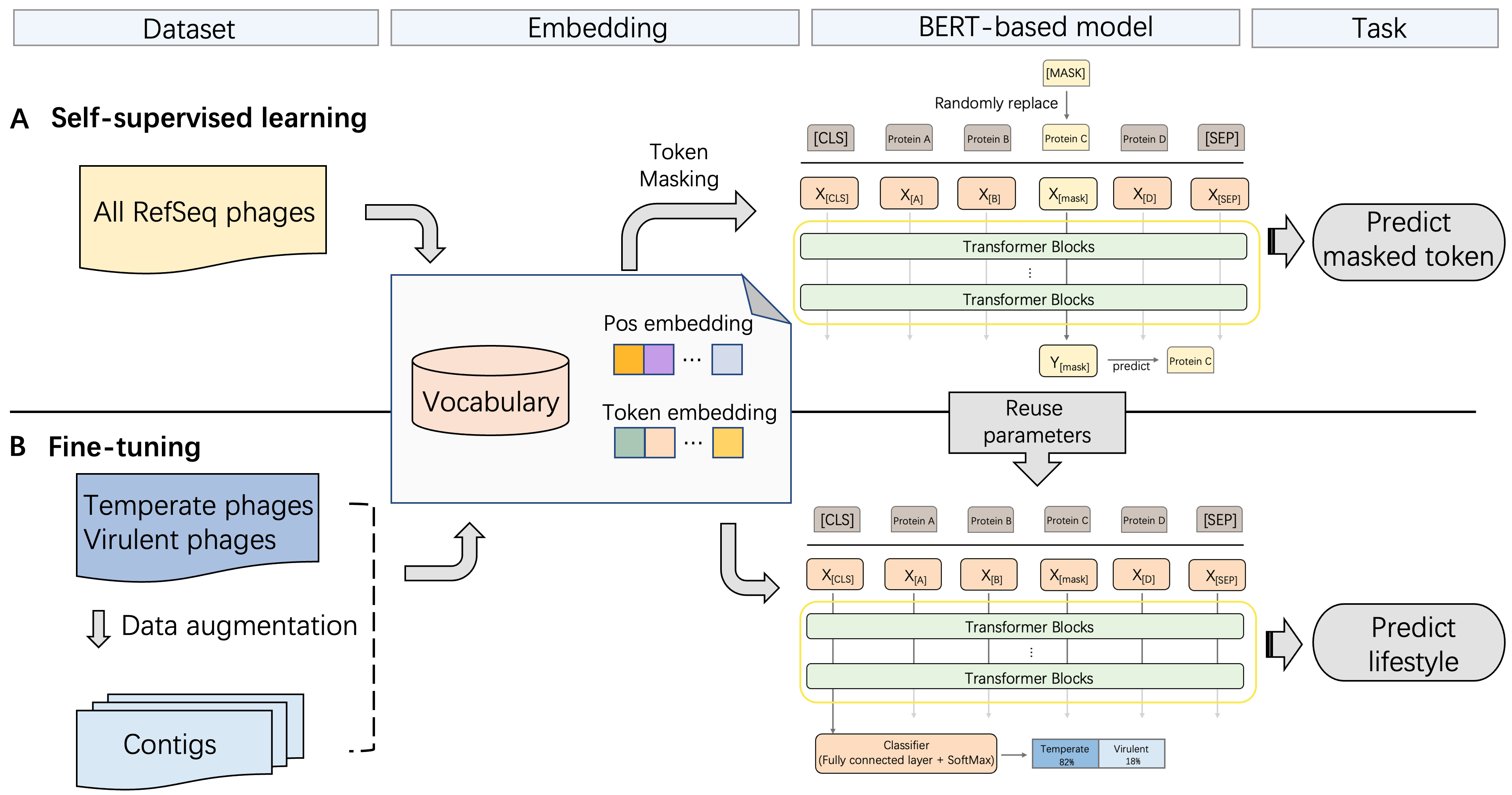}
    \caption{Two training tasks for PhaTYP using BERT. A: the self-supervised learning task. The input is the masked sentence and the output is the predicted token at the masked position. All phage genomes in RefSeq database to train a Mask LM model. B: the fine-tuning task for lifestyle prediction. The pre-trained model is fine-tuned using phages with known lifestyle annotations. The inputs of the model are protein-based sentences and the outputs are the probabilities of two lifestyle classes: virulent and temperate.}
    \label{fig:pipelines}
\end{figure*}

\subsection{Overview}
In this work, we present a method named PhaTYP to classify the lifestyles of phages. Previous works have shown that the marker genes and protein-protein associations are key features in phage classification and identification \cite{chibani2019classifying, pons2021vpf, bin2019taxonomic, shang2021bacteriophage, shang2022cherry, shang2022accurate}. The previous studies also showed that the protein composition and their associations play important roles on phages' lifestyle \cite{pfister1998molecular,emerson2012dynamic}. Inspired by these studies, we represent phage contigs using a contextualized embedding model from natural language processing (NLP) for lifestyle classification. Specifically, we adopt Bidirectional Encoder Representations from Transformer (BERT) to learn the protein composition and associations from phage genomes. We evaluated our final model PhaTYP on contigs of different lengths and contigs assembled from real metagenomic data. The benchmark results against the state-of-the-art methods show that PhaTYP not only achieves the highest performance on complete genomes but also improves the accuracy on short contigs by over 10\%.

\section{Method}
Transformer has emerged as a powerful general-purpose model architecture for representation learning. Because the multi-head mechanism implemented in Transformer can learn the association between tokens, Transformer can be used to generate the semantic representation of the sentences. It outperforms recurrent and convolutional neural networks in several NLP tasks. Inspired by the usage of Transformer in NLP, we propose PhaTYP, which is an eight-layer bidirectional Transformer block based on the original implementation described in \cite{vaswani2017attention}. In NLP problems, words are the tokens in sentences. We make an analogy between words in NLP and proteins in phage genomes so that we can utilize Transformer to learn protein composition and associations from phage genomes. Although k-mers and motifs have been used as tokens for protein prediction task \cite{nambiar2020transforming, du2021secproct}, using proteins as tokens can integrate their biological functions into the sentence representation. In addition, the converted sentence is much shorter by using protein-based tokens, making the training much faster with less parameters. Thus, we train PhaTYP on protein-based tokens to separate virulent and temperate phages.

To address the difficulties of classifying incomplete genomes with limited training data, we divide the lifestyle classification into two tasks: a self-supervised learning task (Fig. \ref{fig:pipelines} A) and a fine-tuning task (Fig. \ref{fig:pipelines} B). In the first task, to circumvent the problem that only a limited number of phages have lifestyle annotations, we applied self-supervised learning to learn protein association features from all the phage genomes using Masked Language Model (Masked LM), aiming to recover the original protein from the masked protein sentences. This task allows us to utilize all the phage genomes for training regardless of available lifestyle annotations. In the second task, we will fine-tune the Masked LM on phages with known lifestyle annotations for classification. To ensure that the model can handle short contigs, we apply data augmentation by generating fragments ranging from 100bp to 10,000bp for training.

In the following section, we will first describe how to convert DNA sequences into protein-based sentences. Then, we will briefly introduce the background of eight-layer bidirectional Transformer blocks. Because we adopt a very standard implementation of Transformer, we refer readers to \cite{vaswani2017attention} for detailed explanation of the model. Finally, we will show how we train PhaTYP on two different tasks: self-supervised training for Masked LM and the fine-tuning model for lifestyle classification.

\subsection{Sequence embedding}

In order to convert sequences into protein-based sentences, we will first introduce how we construct the token vocabulary. Each token in our model represents a protein cluster containing homologous protein sequences from phages. First, we downloaded all the phages proteins from the RefSeq database. We run an all-against-all similarity search using DIAMOND BLASTP \cite{buchfink2015fast} and generate a protein-similarity graph, where the nodes in the graph represent the protein, and the edges connect proteins with significant sequence similarities. The edge weights are the e-values returned from DIAMOND BLASTP. Then, we employ Markov clustering algorithm \cite{enright2002efficient} to group proteins into clusters based on their similarity (e-value). This process resulted in 63,855 protein clusters for constructing the token vocabulary.

\begin{figure}[h!]
    \centering
    \includegraphics[width=0.65\linewidth]{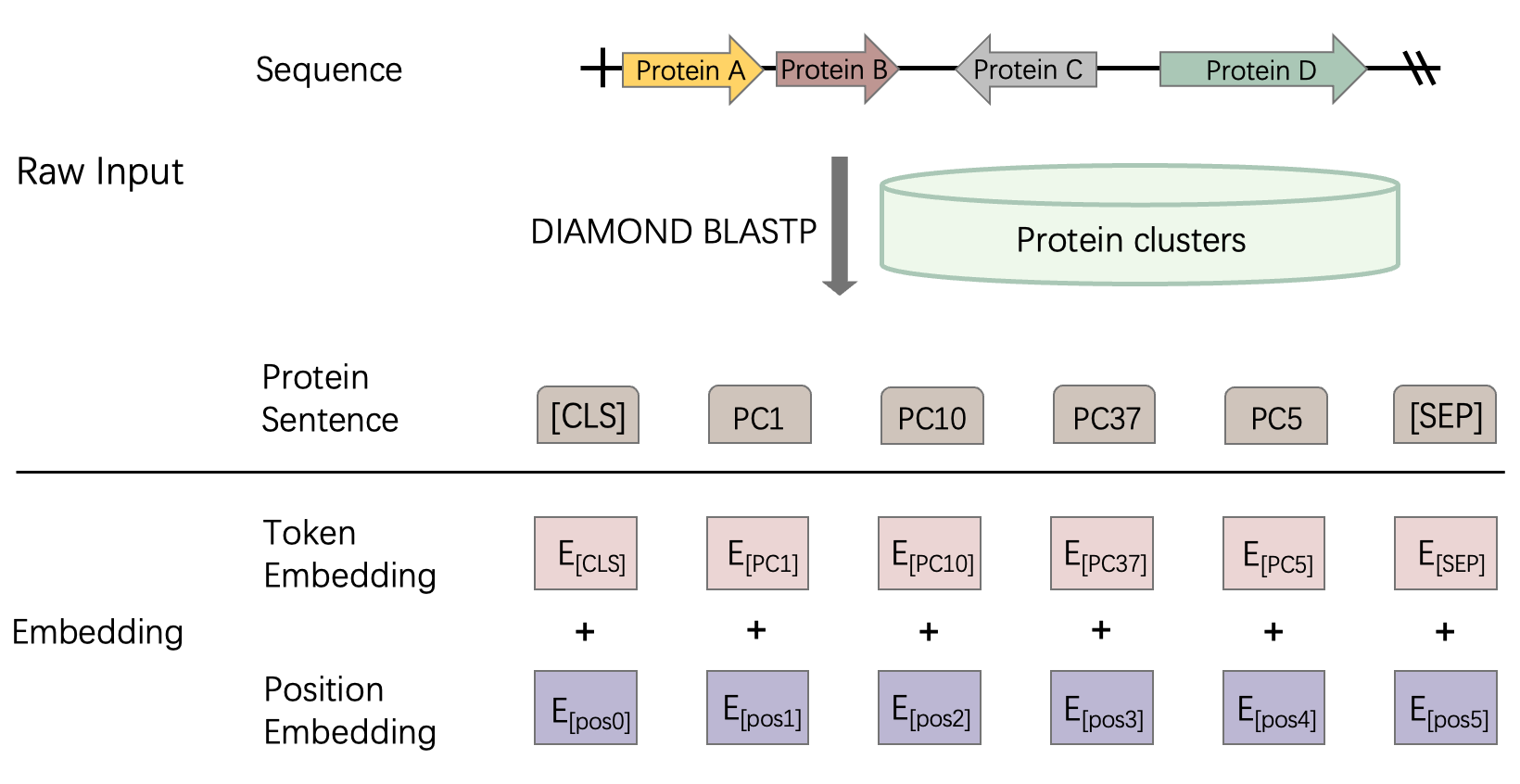}
    \caption{Sequence embedding method in PhaTYP. The block in ``Protein Sentence'' represents the ID of the protein-based token. PC$x$: a protein cluster $x$. [CLS]: start token. [SEP]: separation token. E$_{[PCx]}$: the embedded vector for protein cluster PC$x$. "+": vector addition}
    \label{fig:embedding}
\end{figure}

We then use the generated vocabulary to convert contigs into protein-based sentences. As shown in Fig. \ref{fig:embedding}, first, Prodigal \cite{hyatt2010prodigal} is adopted for gene finding and translation. Then, DIAMOND BLASTP is applied to measure the similarity between these query proteins and the protein clusters. Each query protein is assigned with the token with the best alignment. Finally, the contigs can be converted into protein-based sentences as shown in Fig. \ref{fig:embedding}. Because the length of the sentences can vary a lot, we follow the design of BERT \cite{devlin2018bert} and choose 300 as the maximum length of the sentences. Thus, the sentence is a 300-dimensional vector, with each dimension encoding a token ID. Two specialized tokens, [\textit{CLS}] and [\textit{SEP}], representing the start of the sentences and separation of the sentences, are the first and last tokens in each sentence. If the contigs have more than 298 tokens, we only keep the first 298 tokens. On the contrary, if the contigs contain less than 298 tokens, we pad token [\textit{PAD}] at the end of the sentences.

After converting the DNA sequences into protein-based sentences, we will project the 300-dimensional vector into a dense embedding matrix. We employ a learnable embedding layer, which is a neural network, to embed the sentence. There are two main purposes of using the learnable embedding layer. First, the learnable embedding layer will generate a low-dimensional embedding vector than using one-hot encoding as input, which can result in resulting in a $\mathbb{R}^{300 \times 63,855}$ matrix for each sentence. The deep learning model will suffer from the curse of dimensionality using such a sparse matrix as input \cite{mikolov2013distributed}. Second, As proven in \cite{mikolov2013distributed}, the embedding layer can learn to map associated tokens into similar embedding vectors, and thus assisting the learning process. In addition, as shown in Fig. \ref{fig:embedding}, we embed the position information to represent the position of the token in the sentence, helping the model utilize the sequential information of the sentence. The equations of the embedding layers are listed in Eqn. \ref{Eq1}.

\begin{equation}\label{Eq1}
\left\{\begin{matrix}
\widetilde{E_t} = Embed(E_t, W_{E_{t}})\\
\widetilde{E_p} = Embed(E_p, W_{E_{p}})\\
X = \widetilde{E_s} + \widetilde{E_p}
\end{matrix}\right.
\end{equation}


\noindent $E_t \in \mathbb{R}^{300 \times 1}$ represents the token ID sentence and $E_p \in \mathbb{R}^{300 \times 1}$ represents the position index vector. $W_{E_{t}} \in \mathbb{R}^{63,855 \times 512}$ and $W_{E_{p}}  \in \mathbb{R}^{300 \times 512}$ are the learnable projection matrices. 512 is the default embedding dimension suggested by \cite{vaswani2017attention}. The function of $Embed$ is a look-up table. Given an ID, it returns the corresponding embedding vector. 
Thus, the dimension of $\widetilde{E_t}$ and $\widetilde{E_p}$ are $\mathbb{R}^{300 \times 512}$, which is much smaller than using the one-hot encoding matrix ($\mathbb{R}^{300 \times 63,855}$). We follow the design of BERT \cite{devlin2018bert} and train the embedding layer with the whole model via the end-to-end mode to enlarge the receptive field. Finally, we will apply matrix addition on the two embedded matrices (defined as X in Eqn. 1) and feed it into the eight-layer Transformer structure.

\subsection{Model structure}

\begin{figure}[h!]
    \centering
    \includegraphics[width=0.35\linewidth]{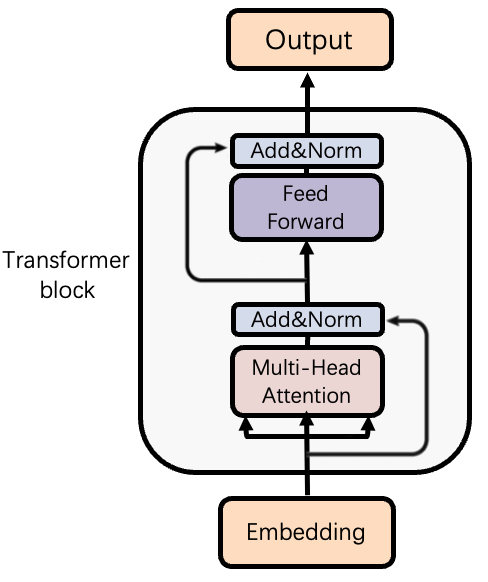}
    \caption{The Transformer block in PhaTYP. There are three main units in the transformer block: feed-forward network, residual connections, and multi-head attention mechanism.}
    \label{fig:Transformer}
\end{figure}

The detailed architecture of the Transformer model is shown in Fig. \ref{fig:Transformer}. The main function is the multi-head attention mechanism, which can extract association features between tokens \cite{vaswani2017attention}. The feed-forward network consists of two fully connected layers with a ReLU activation, which is applied to each position separately and identically. To prevent overfitting, residual connections \cite{he2016deep}, and layer normalization \cite{ba2016layer} are employed before and after the feed-forward network. The input of the Transformer block is the embedding $X$ in Eqn. \ref{Eq1}. The self-attention mechanism and feed-forward network of the Transformer grant access to all protein tokens in the embedded sentence and leverage the association information between tokens to generate latent feature output $Y$. Because of the hyper-parameter setting in the Transformer model, $Y$ has the same dimension as the embedding $X$. In PhaTYP, we stack eight Transformers in serial. In each Transformer, we employ eight-head attention to focus on different representation sub-spaces. 

After introducing the model structure, we will show how to train PhaTYP on Masked LM and fine-tuning tasks. These two tasks have different but related objective functions. In the following sections, we will first describe the datasets used in each task and then present a detailed explanation of the loss function.

\subsection{Self-supervised training for Masked LM}
Self-supervised learning is a kind of unsupervised learning aiming to learn a general feature expression for downstream tasks. In our design, the downstream task is the lifestyle classification, and the general feature is the protein organizations in phage genomes. Because the number of known lifestyle annotation of phages is far less than the number of known phages in the database, we employ a self-supervised learning method to learn as much prior knowledge as possible. Specifically, we use all the available phage genomes in RefSeq to train the Masked LM task. 

\paragraph{Dataset}
We download all the phage genomes released before 2022 from the NCBI RefSeq database. In total, we have 3,474 phage genomes. To generate more data for training, we cut the complete genomes into fragments with different lengths, including 5kbp, 10kbp, 15kbp, and 20kbp. We randomly sample ten sub-strings from the genomes for each length. Thus, we have 142,434 phage contigs in our dataset. Finally, we will use the method introduced in section \textit{sequence embedding} to generate protein sentences and embedding as inputs to Transformer.

\paragraph{Masked LM loss}
The inputs to our self-supervised model are masked sentences, and the aim is to predict the original sentences. For example, as shown in Fig. \ref{fig:pipelines} A, we replace the embedding at the position of \textit{protein C} with the [\textit{MASK}] token. Then, we will feed the masked sentence into PhaTYP to predict the marked token.

As described in Section \textit{model structure}, the shape of output $Y$ will be the same as the input $X$, which is $\mathbb{R}^{300 \times 512}$. Thus, each row $i$ in $Y$ can be interpreted as the latent features of the input token at position $i$ in the sentence, which is a $\mathbb{R}^{1 \times 512}$ vector. Then, in order to predict the ID of the masked token, $Y_i$ is fed into an output layer with SoftMax function over the vocabulary. The loss function is presented in Eqn. \ref{Eq:loss1}.

\begin{equation}\label{Eq:loss1}
loss(SoftMax(Y_iW_d), Token_i)
\end{equation}

\noindent $W_d \in \mathbb{R}^{512 \times 63,855}$ is the learnable projections. The value in $SoftMax(Y_iW_d) \in \mathbb{R}^{1 \times 63,855}$ represents the prediction probability of each token in the vocabulary. In the training process, 5\% words of each sentence will be masked randomly, and we will calculate the cross-entropy loss to update the parameters through backpropagation.

\subsection{Fine-tuning the model for lifestyle classification}
After pre-training the model via the self-supervised learning task, we will then fine-tune PhaTYP to the downstream task: lifestyle classification.

\paragraph{dataset}
There are two widely used datasets supplied by the previous studies \cite{mcnair2012phacts, wu2021deephage}. In total, we have 1,290 virulent phages and 577 temperate phages. In order to balance the dataset and improve the robustness on short contigs, data augmentation is applied by randomly generating short DNA fragments, ranging from 100bp to 20kbp, from the complete genomes. For each length, we generate 10,000 contigs for temperate and virulent phages, resulting in 160,000 phage contigs in the dataset. Then, all the fragments will be converted into protein sentences and fed into PhaTYP.

\paragraph{Fine-tuning loss}
Unlike the self-supervised learning task, we use the first row $Y_0 \in \mathbb{R}^{1 \times 512}$ from output $Y$ as latent feature for the lifestyle classification. As shown in Fig. \ref{fig:pipelines} B, we concatenate a fully connected layer to calculate the probability of the input being virulent or temperate. The loss function of the fine-tuning can be found in Eqn. \ref{Eq:loss2}

\begin{equation}\label{Eq:loss2}
loss(SoftMax(Y_0W_p+b_p), label)
\end{equation}

\noindent $W_d \in \mathbb{R}^{512 \times 2}$ and $b_p \in \mathbb{R}^{1 \times 2}$ are the learnable parameters. Then, we will calculate the loss between the real labels and the predictions and update the parameters in the model accordingly.

\subsection{Model training}
In the training process, we employ ten-fold cross-validation for both tasks. The model is trained with 4 GTX 2080Ti GPU units using the Adam optimizer, and we apply a learning rate of 0.001 to update the parameters. Because PhaTYP needs to be trained step-by-step on the two tasks, we first choose the model with the lowest cross-validation loss on self-supervised learning tasks. Then, we fine-tune this pre-trained model on the lifestyle classification task. Finally, we store the parameters of the model that achieve the best performance (AUCROC) in the cross-validation. Both the parameters and the model are available via \url{http://github/KennthShang/PhaTYP/}.

\section{Result}
In this section, we validate PhaTYPE on both simulated and real sequencing data. We compared PhaTYP against the state-of-the-art methods, including PHACTS \cite{mcnair2012phacts}, DeePhage \cite{wu2021deephage}, BACPHLIP \cite{hockenberry2021bacphlip}, and PhagePred \cite{song2020classifying}. Because the authors of BACPHLIP stated that incomplete/partially assembled genomes should not be used as input, we will only evaluate BACPHLIP on complete genomes. It is worth noting that PhagePred and DeePhage allow re-training using customized training data. Thus we re-trained these models with the same training set as PhaTYP for all experiments. PHACTS and BACPHLIP did not provide this function and thus, we used the provided model in all experiments. 

\paragraph{Metrics} To ensure consistency and a fair comparison, we use the same metrics as the previous works: sensitivity ($TP/(TP+FN)$), specificity ($TN/(TN+FP)$), and accuracy ($(TP+TN)/(TP+FN+TN+FP)$). $TP$ represents the number of correctly classified temperate phages, while $TN$ represents the number of correctly classified virulent phages. Thus, sensitivity and specificity can be interpreted as the recall of classifying temperate and virulent viruses, respectively. In addition, we will present the ROC curve to evaluate the tradeoffs between sensitivity and specificity. 

\begin{table}[h!]
\centering

\begin{tabular}{p{7cm}p{8cm}}
\hline
Name                                   & Description                                                                                                                                                                            \\ \hline
RefSeq                                 & All the phage genomes released before 2022 from the NCBI RefSeq database. Totally, we have 3,474 phage genomes. This dataset is only used to train the Masked LM task. \\
Lifestyle dataset 1                    & Lifestyle annotations from the dataset of \cite{mcnair2012phacts}, including 77 virulent phages and 148 temperate phages.                                             \\
Lifestyle dataset 2                    & Lifestyle annotations from the dataset of \cite{wu2021deephage}, including 1,211 virulent phages and 429 temperate phages.                                            \\
Phage contigs from human neonates' guts & 2,291 phage contigs assembled from metagenomic data. The metagenomic data is sampled from 20 human neonates' guts \cite{liang2020stepwise}.    \\ \hline                        
\end{tabular}
\caption{Detailed information of the dataset used in the experiments.}
\label{tab:data}
\end{table}

\paragraph{Datasets} As described  in Section \textit{Self-supervised training for Masked LM} and \textit{Fine-tuning the model for lifestyle classification}, totally we have three datasets for training and validation. In addition, we applied PhaTYP to predict the lifestyle for phage contigs assembled from human neonates' guts. The detailed information of the datasets is listed in Table \ref{tab:data}.

\subsection{Classification performance comparison using ten-fold cross-validation}
\paragraph{Performance on the complete genomes}
We applied ten-fold cross-validation on the combined lifestyle datasets listed in Table \ref{tab:data}. When training, we apply the data augmentation method mentioned in section \textit{Fine-tuning the model for lifestyle classification} on the training set. The phage sequences in the validation set remain complete. 

\begin{figure}[h!]
    \centering
    \includegraphics[width=0.55\linewidth]{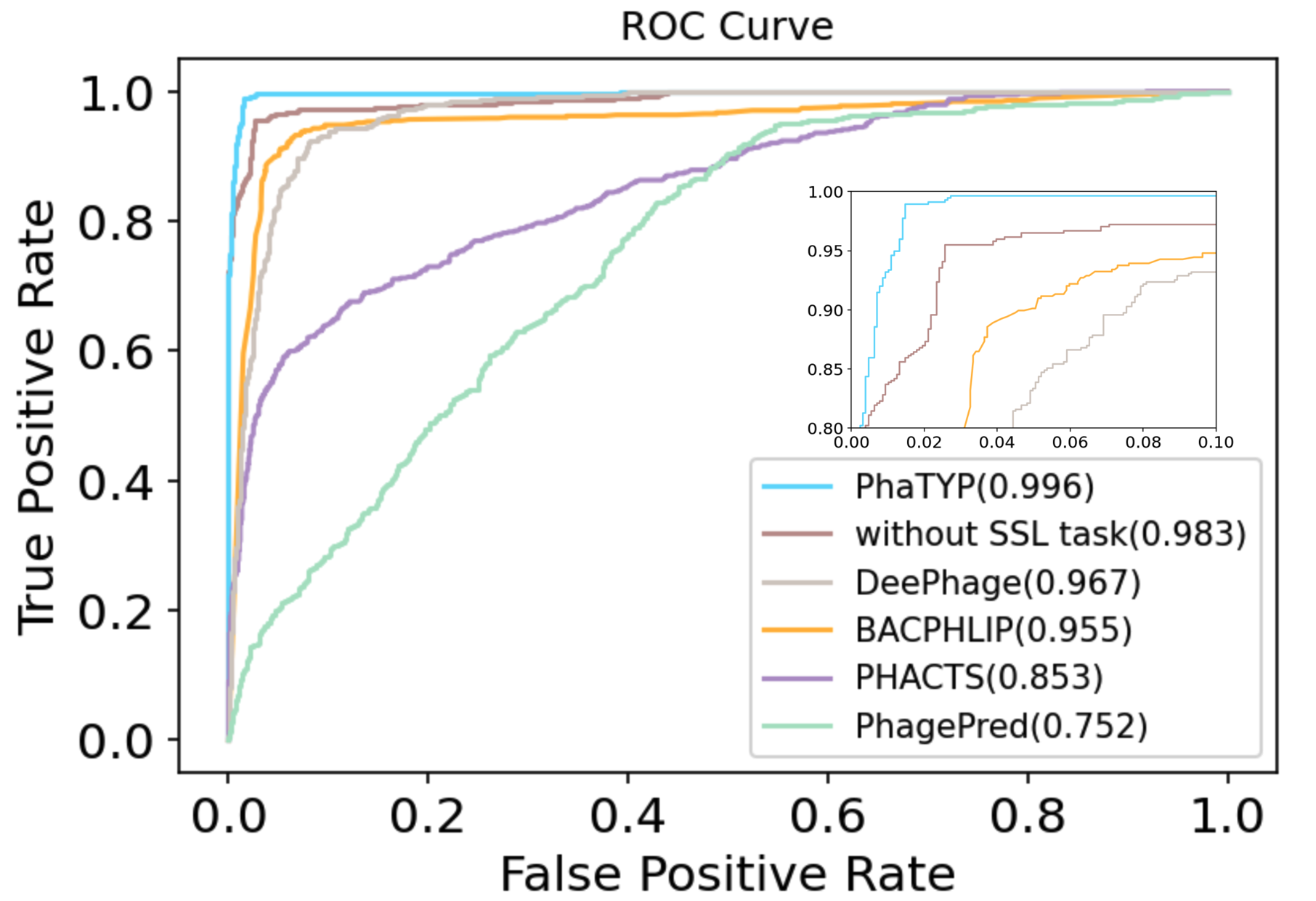}
    \caption{The ROC curve comparison on the complete phage genomes. The value shown in the legend is the AUCROC score. ``without SSL task'': training PhaTYP without the self-supervised learning task.}
    \label{fig:complete}
\end{figure}

The ROC curves derived from the averaged ten folds evaluation from all the methods are shown in Fig. \ref{fig:complete}. We also show how the self-supervised learning task affects the learning performance by recording the results of training PhaTYP without the self-supervised learning task (\textit{without SSL task} in Fig. \ref{fig:complete}). Predicting the lifestyle for complete phages genomes is a relatively easy task. PhaTYP, DeePhage, and BACPHLIP all achieved high accuracy. Nevertheless, the ROC curves in Fig. \ref{fig:complete} show that PhaTYP has the best performance. 

\begin{table}[h!]
\centering
\begin{tabular}{cccc}
\hline
\textbf{Metrics}          & \textbf{sensitivity} & \textbf{specificity} & \textbf{Acc}  \\ \hline
\textbf{PhaTYP}           & \textbf{0.99}          & \textbf{0.97}          & \textbf{0.98} \\
\textbf{without SSL task} & 0.97                   & 0.91                   & 0.95          \\
\textbf{DeePhage}         & 0.94                   & 0.93                   & 0.94          \\
\textbf{BACPHLIP}         & 0.97                   & 0.87                   & 0.93          \\
\textbf{PHACTS}           & 0.91                   & 0.75                   & 0.85          \\
\textbf{PhagePred}        & 0.69                   & 0.88                   & 0.81          \\ \hline
\end{tabular}
\caption{Detailed results on the complete genomes under each tools' default score cutoffs (0.5).}
\label{tab:complete}
\end{table}

In addition, we listed more detailed information about the performance of predicting virulent and temperate phages in Table \ref{tab:complete}. The results reveal that PhaTYP can identify temperate phages with higher specificity than other tools. The ablation study also shows that training with the self-supervised learning task can improve classification accuracy.

\begin{table}[h!]
\centering
\begin{tabular}{lllllll} \hline
Program                       & PhaTYP &  DeePhage & BACPHLIP & PHACTS & PhagePred \\ \hline
Elapsed time(min) & 21     & 3      & 55     & 87   & 446     &  \\ \hline
\end{tabular}
\caption{The elapsed time to make predictions for the ten-fold cross-validation. All the methods are run on Intel\textsuperscript{\textregistered} Xeon\textsuperscript{\textregistered} Gold 6258R CPU and 2080Ti GPU.}
\label{tab:time}
\end{table}

we also recorded the running time of the five methods on the ten-fold cross-validation in Table \ref{tab:time}. PhaTYP is not the fastest tool because $\sim$85\% running time is used to run DIAMOND BLASTP as described in Section \textit{Sequence embedding}. DeePhage requires less running time because it only uses $k$-mer features, while other methods also incorporate alignment features for lifestyle prediction.

\paragraph{Performance on the test set with low similiairty }
Cross validation cannot control the similarity between the training and test set. To evaluate whether the learning models can perform well on a harder test set, we constructed a test set containing phage genomes with low similarity with the training set. First, we applied Dashing \cite{baker2019dashing} to estimate the similarity between phages in the lifestyle dataset. 
Then, we generated a test set by selecting 50 phages with the lowest similarities with their peers from virulent and temperate phages, respectively. In total, there are 100 phages in the test set. The maximum similarity of the virulent phages and temperate phages in the test set with all other phages are 0.049 and 0.052, respectively.

\begin{figure}[h!]
    \centering
    \includegraphics[width=0.55\linewidth]{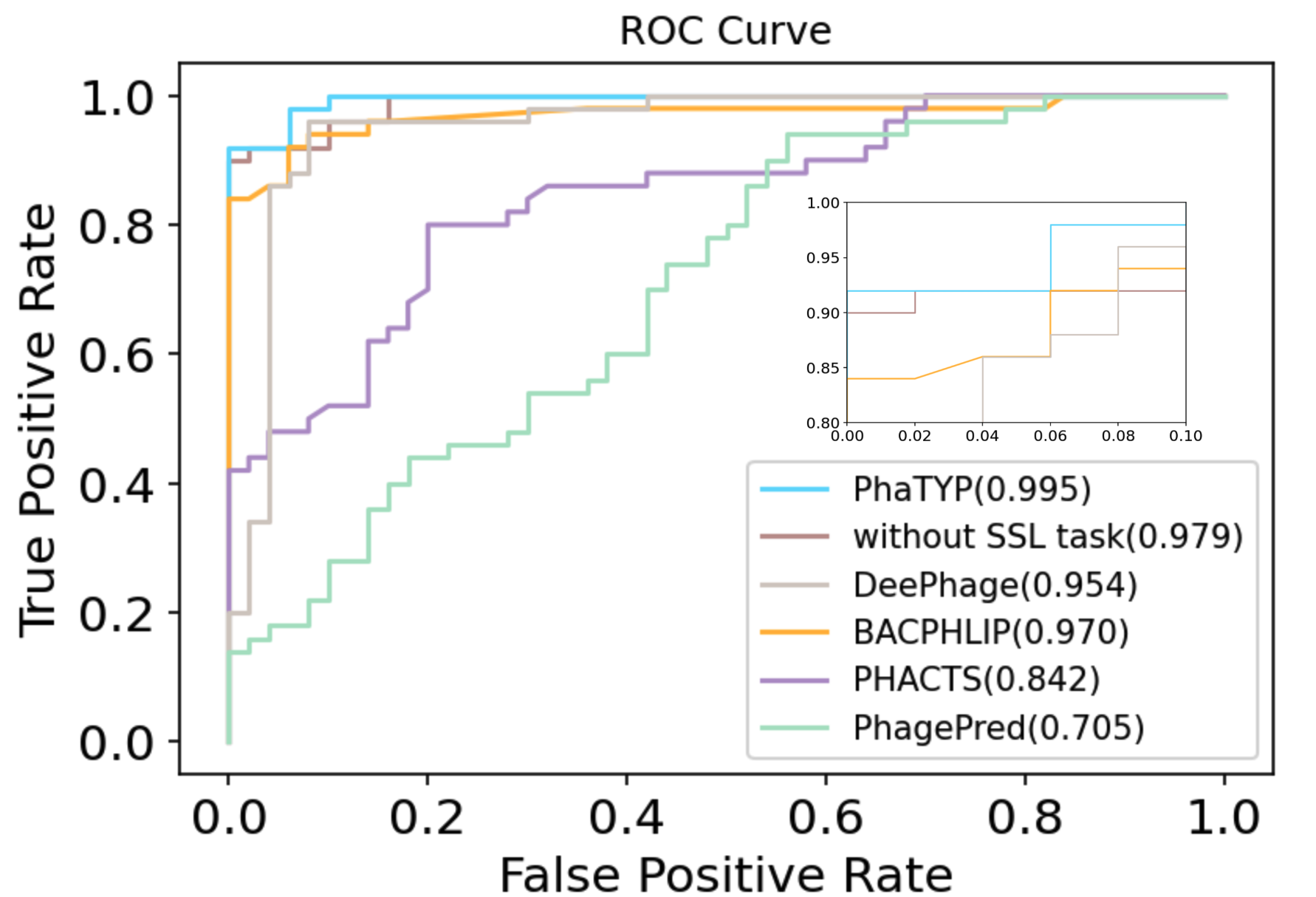}
    \caption{The ROC curve comparison on low similarity test set. The value shown in the legend is AUCROC score. ``without SSL task'': training PhaTYP without self-supervised learning task.}
    \label{fig:similarity}
\end{figure}

\begin{table}[h!]
\centering
\begin{tabular}{cccc}
\hline
\textbf{Metrics}          & \textbf{sensitivity} & \textbf{specificity} & \textbf{Acc} \\ \hline
\textbf{PhaTYP}           & \textbf{0.99}          & \textbf{0.89}         & \textbf{0.94}        \\
\textbf{without SSL task} & 0.98                   & 0.86                  & 0.92         \\
\textbf{DeePhage}         & 0.96                   & 0.86                  & 0.91         \\
\textbf{BACPHLIP}         & 0.98                   & 0.84                  & 0.90         \\
\textbf{PHACTS}           & 0.90                   & 0.69                  & 0.74         \\
\textbf{PhagePred}        & 0.57                   & 0.83                  & 0.67         \\ \hline
\end{tabular}
\caption{The performance comparison on the low similarity test set under the tools' default score cutoff (0.5).}
\label{tab:similarity}
\end{table}

The classification results are shown in Table \ref{tab:similarity} and Fig. \ref{fig:similarity}. Compared to the cross-validation results in Table \ref{tab:complete}, the similarity of the test genomes affects specificity more than sensitivity, indicating that some virulent phages are misclassified as temperate. This could be caused by lower average similarity for the test virulent phages. In Fig. \ref{fig:similarity}, the value of AUCROC decreases for all methods except BACPHLIP. This is likely caused by the overlap between the test phages and the data used for training BACPHLIP in its provided version. Nevertheless, PhaTYP still has the best results.

\paragraph{Performance on the short contigs}
Considering that metagenomic assembly can produce incomplete phage sequences, it is important to evaluate PhaTYP on phage contigs. Towards this goal, we apply the same method used in \cite{wu2021deephage} to construct the short fragment dataset. Specifically, we generate four groups of fragments with different length ranges, including 100-400bp, 400-800bp, 800bp-1,200bp, and 1,200-1,800bp. These four sets of fragments can cover the length of raw reads and the short contigs from the metagenomic data. We generated 80,000 fragments for each group by random sampling a sub-string from the complete genomes, with 40,000 for temperate and virulent phages, respectively. To remove the potential redundant fragments, we used Dashing \cite{baker2019dashing} to estimate the similarity between fragments. Then, we removed fragments with similarities above 0.8 from the dataset. We applied ten-fold cross-validation to train and validate the performance. Because DeePhage trained four separate models on different length ranges, we followed the same design and trained four PhaTYP models separately.

\begin{figure}[h!]
    \centering
    \includegraphics[width=0.55\linewidth]{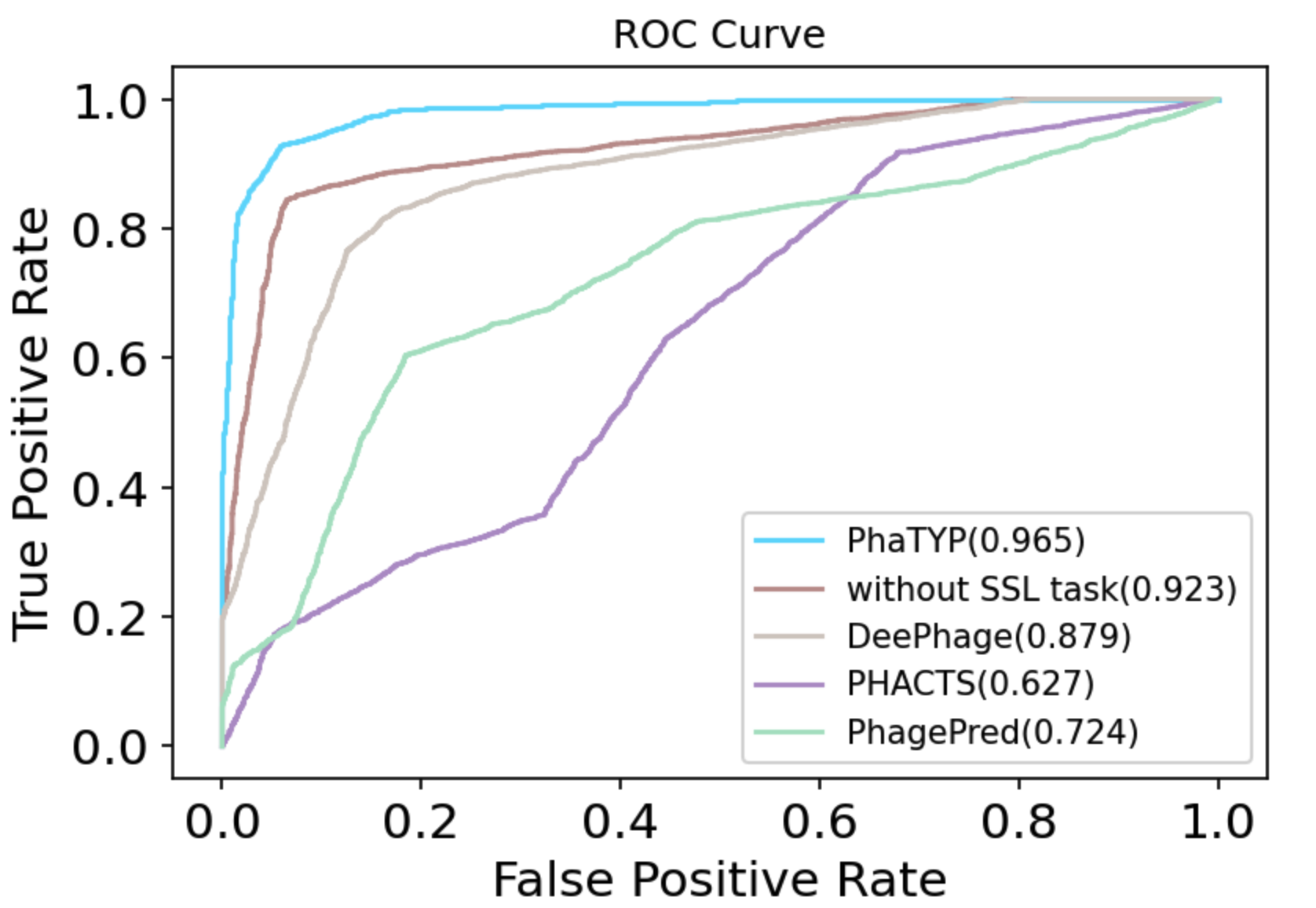}
    \caption{The ROC curve comparison on the short contigs. The value shown in the legend is AUCROC score. ``without SSL task'': training PhaTYP without self-supervised learning task. PhaTYP has the best performance.}
    \label{fig:shortcontigs}
\end{figure}

\begin{table}[h!]
\centering
\begin{tabular}{cccccc}
\hline
\textbf{Tool}                                               & \textbf{Criterion} & \textbf{100-400} & \textbf{400-800} & \textbf{800-1200} & \textbf{1200-1800} \\ \hline
                                   & sensitivity                 & \textbf{0.86}    & 0.91    & \textbf{0.93}     & 0.93      \\
                                 & specificity                 & \textbf{0.92}             & 0.92             & \textbf{0.93}              & \textbf{0.94}      \\

\multirow{-3}{*}{\textbf{PhaTYP}}   & Accuracy                & \textbf{0.89}             & 0.91             & \textbf{0.93}              & \textbf{0.94}      \\
                                                            & sensitivity                 & 0.83             & 0.85             & 0.87              & 0.92      \\
                                                            & specificity                 & 0.92             & \textbf{0.93}             & 0.92              & 0.93      \\
\multirow{-3}{*}{\textbf{without SLL task}}                 & Accuracy                & 0.87             & 0.89             & 0.90              & 0.92      \\
                                   & sensitivity                 & 0.74             & 0.84             & 0.86              & 0.89      \\
                                   & specificity                 & 0.77             & 0.82             & 0.86              & 0.87      \\

\multirow{-3}{*}{\textbf{DeePhage}} & Accuracy                & 0.76             & 0.83             & 0.86              & 0.88      \\
                                                            & sensitivity                 & 0.56             & 0.60             & 0.62              & 0.64      \\
                                                            & specificity                 & 0.75             & 0.80             & 0.84              & 0.87      \\
\multirow{-3}{*}{\textbf{PhagePred}}                                 & Accuracy                & 0.65             & 0.70             & 0.72              & 0.75      \\
                                  & sensitivity                 & 0.26             & 0.36             & 0.39              & 0.42      \\

                                  & specificity                 & 0.73             & 0.64             & 0.65              & 0.69      \\

\multirow{-3}{*}{\textbf{PHACTS}}            & Accuracy                & 0.48             & 0.49             & 0.51              & 0.54      \\ 
                                                            & sensitivity                 & \textbf{0.86}             & \textbf{0.92}             & 0.92              & \textbf{0.94}      \\
                                                            & specificity                 & 0.90             & 0.92             & \textbf{0.93}              & \textbf{0.94}      \\
\multirow{-3}{*}{\textbf{PhaTYP (augmentation)}}                                 & Accuracy                & 0.88             & \textbf{0.92}             & \textbf{0.93}              & \textbf{0.94}      \\
\hline
\end{tabular}
\caption{Detailed results on the short contigs under the tools' default score cutoff (0.5).}
\label{tab:shortcontigs}
\end{table}

We generate the ROC curve on short contigs to show the tradeoff between FP rate and sensitivity. In each fold, we combine the validation results from the four length ranges. The results are shown in Fig. \ref{fig:shortcontigs}, which shows that our model outperforms the other tools. More detailed results are shown in Table \ref{tab:shortcontigs}. In general, with the increase of the length, the performance of all methods increases. This is expected because longer fragments generally provide more information for prediction. In addition, compared to the results on complete genomes, the self-supervised learning task can help the model achieve higher accuracy on short contigs. A plausible explanation is that pre-training on the self-supervised learning task can help the model learn more generalized embeddings for phage proteins and prevent overfitting. Then, these embeddings with prior knowledge can be leveraged when information for classification is lacking. In addition, we tested whether training contigs with different lengths at the same time influences the performance of PhaTYP. Specifically, we employed the data augmentation methods, which combines the training set of the contigs and their complete genomes to train one PhaTYP model. The comparison between training PhaTYP separately and training with all data at once (PhaTYP augmentation) in Table \ref{tab:shortcontigs} reveals that training with data augmentation will not affect the classification performance. Thus, PhaTYP can predict lifestyles for contigs across different length ranges with one model, which helps save computational resources for users.

\subsection{Predicting the lifestyle of crAssphages}
CrAssphages are an extensive family of tailed bacteriophages discovered through the cross-assembly of human fecal metagenomes \cite{dutilh2014highly}. Most of them infect Bacteroides and do not integrate into their hosts during replication. Thus, these crAssphages are widely regarded as virulent phages. Despite its ubiquity in human gut, over 80\% of the predicted proteins in crAssphage genomes showed no significant similarity to annotated crAssphage protein sequences, hampering their identifications in newly sequenced metagenomic data \cite{yutin2018discovery}. The low similarity also poses a hard case for current lifestyle classification tools. 

In this experiment, we downloaded 33 crAssphages from \cite{liang2020stepwise}, which are all annotated as virulent phages, and evaluated whether PhaTYP can correctly classify them. First, we remove all the training sequences with high sequence similarity (alignment identity $>50$\%) from our dataset using BLASTN \cite{camacho2009blast+}. Then, we retrained PhaTYP, DeePhage, and PhagePred on the `cleaned' dataset.

\begin{figure}[h!]
    \centering
    \includegraphics[width=0.35\linewidth]{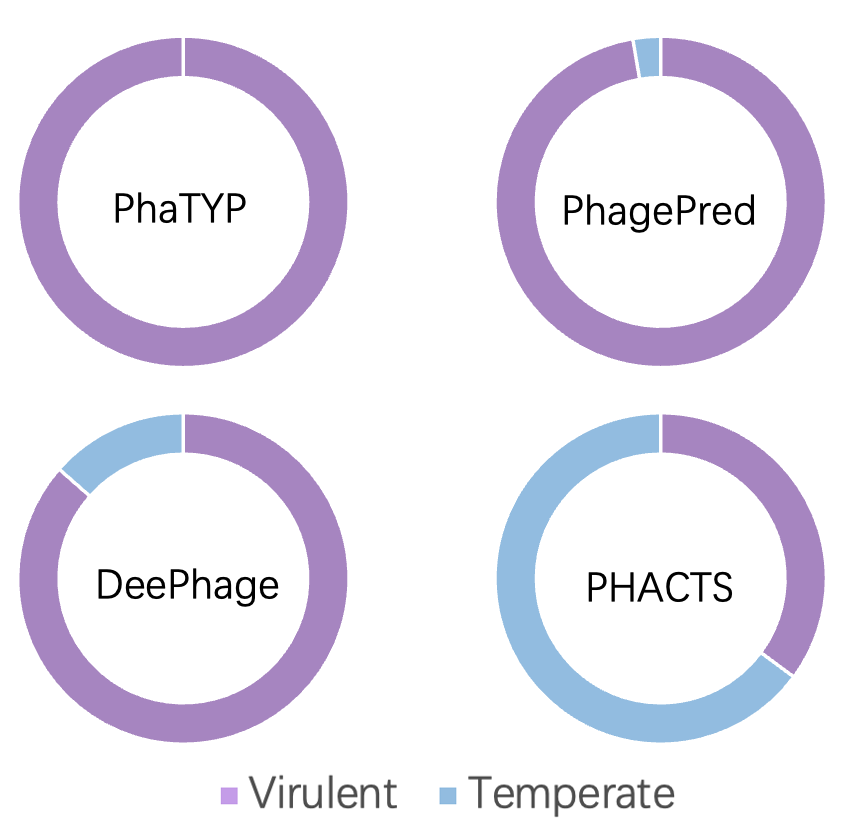}
    \caption{The classification results on the 33 virulent crAssphages. PhaTYP can correctly predict all the crAssphages as being virulent.}
    \label{fig:crassphage}
\end{figure}

The results of this experiment are presented in Fig. 7. All the 33 crAssphages can be classified correctly by PhaTYP, further demonstrating its utility. PhagePred ranked the second, showing much improved performance than in the previous experiments. One possible reason is that the k-mer frequency feature adopted by PhagePred is an important feature for crAssphages.

\subsection{Case study: phage lifestyle analysis for infants}
In this section, we apply PhaTYP to investigate early-life viral colonization using the metagenomic data sampled from infant meconium or early stools \cite{liang2020stepwise}. In the original study, the authors used reference genomes and PHACTS to analyze the phages' lifestyles. Although the similarity search against the reference genomes can be used to annotate the lifestyles, the number of aligned phages is only a small subset of all assembled phage contigs. Then, the authors used PHACTS to predict lifestyle of all the phage contigs. However, according to the previous experiments, the performance of PHACTS is not ideal on short contigs. Thus, we employed PhaTYP to re-investigate this dataset.

The data is sequenced from 20 infants. The stool samples were collected longitudinally at 0–4 days after birth (meconium samples, \textit{month 0}), \textit{month 1}, and \textit{month 4}. Therefore, we have a total of 60 metagenomic samples. In addition, the authors recorded detailed information, including the feeding and delivery type of these infants. Thus, we are able to investigate whether the ages, feeding, and delivery type can affect the composition of phages with different lifestyles. The 60 samples containing the raw reads are public and available via \url{https://www.ncbi.nlm.nih.gov/sra/?term=PRJNA524703}; the assembled contigs are available via \url{https://github.com/guanxiangliang/liang2019}. 

Following the guidelines of \cite{liang2020stepwise}, we first applied Bowtie2 \cite{langmead2012fast} for reads mapping and then ran BBmap \cite{bushnell2014bbmap} to calculate the reads per million total reads (RPM) for each contig. According to the description in \cite{liang2020stepwise}, the species were called present if at least ten RPM from one sample aligned to that contig. Therefore, we have 19,943 contigs remained. Second, to ensure consistency, we used the same rules to define phage contigs as in \cite{bushnell2014bbmap}: (1) at least one phage protein per 10kb of the contig and (2) 50\% of the predicted open-reading frames (ORFs) are phage ORFs. We used the Prodigal \cite{hyatt2010prodigal} in meta mode to predict the ORFs and aligned them to the phage protein database downloaded from RefSeq. Finally, 2,291 high-confidence  phage-like contigs were identified and we ran PhaTYP for lifestyle prediction. 

\begin{figure}[h!]
    \centering
    \includegraphics[width=0.35\linewidth]{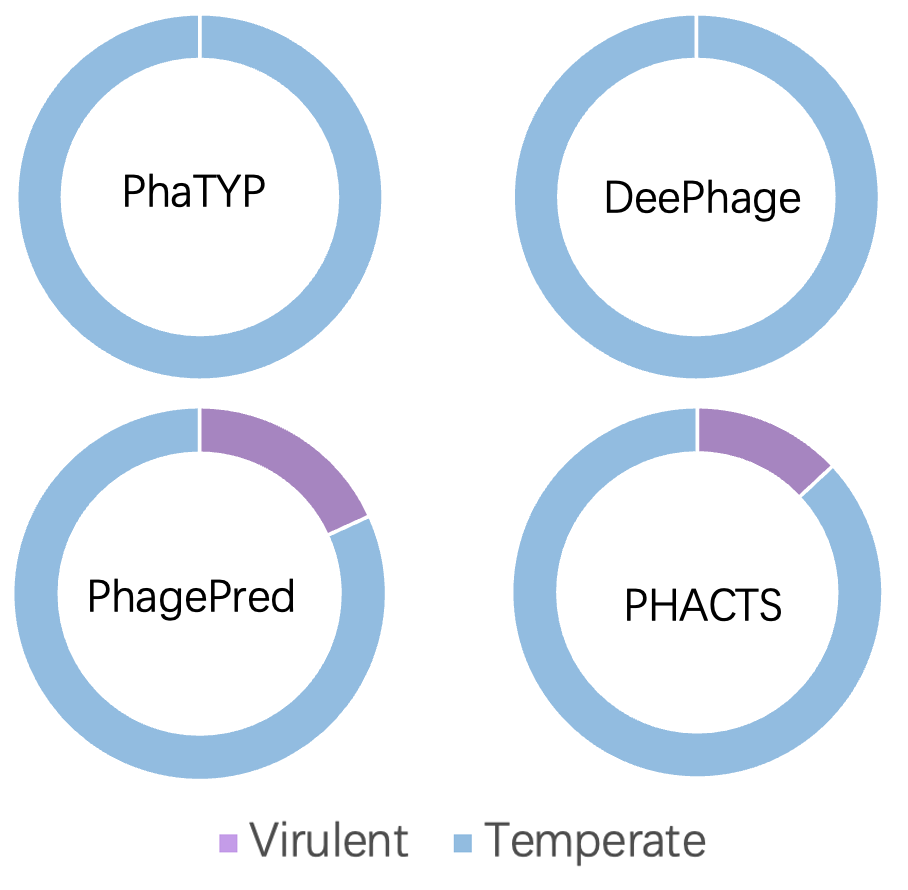}
    \caption{The classification results on the contigs that have homologous regions with integrase proteins. Both PhaTYP and DeePhage can correctly predict all the contigs as being temperate.}
    \label{fig:integrase}
\end{figure}

\paragraph{Identifying temperate phages containing integrase}
Because there are no labels for the contigs from the metagenomic data, we rely on the marker genes for labeling the contigs. As we discussed in section \textit{Introduction}, using marker genes usually has high specificity but low recall. Thus, the contigs labeled using this method have high confidence.  One of the most widely accepted markers, named integrase, is commonly used to identify temperate phages \cite{zink1992classification, denes2014comparative}. Thus, we built an integrase protein database by searching the proteins with the keyword `integrase' from the phage protein database. Then, we search all 2,291 phage contigs against the integrase database using BLASTX \cite{camacho2009blast+}. 23 contigs having homologous regions (identity $>$ 90\%, coverage $>$ 90\%, and e-value $<$ 1e-10) were identified out of 2,291 contigs. All these 23 contigs were labeled as temperate phages, and we used them to evaluate PhaTYP and the state-of-the-art tools.

As shown in Fig. \ref{fig:integrase}, both PhaTYP and DeePhage can predict correctly on all 23 contigs. On the other hand, the results show that using the integrase as a marker can only classify \textasciitilde1\% of the metagenomic phages, which remains a low recall for lifestyle prediction. Thus, methods that do not just rely on integrase proteins are in great need of comprehensive prediction. 

\paragraph{Comprehensive lifestyle analysis}
Then, we analyzed all 2,291 phages' lifestyles using PhaTYP. There are three main variables that might affect the dominant lifestyle: the \textit{age} of the infant, the \textit{delivery} type, and the \textit{feeding} type.

\begin{figure}[h!]
    \centering
    \includegraphics[width=0.65\linewidth]{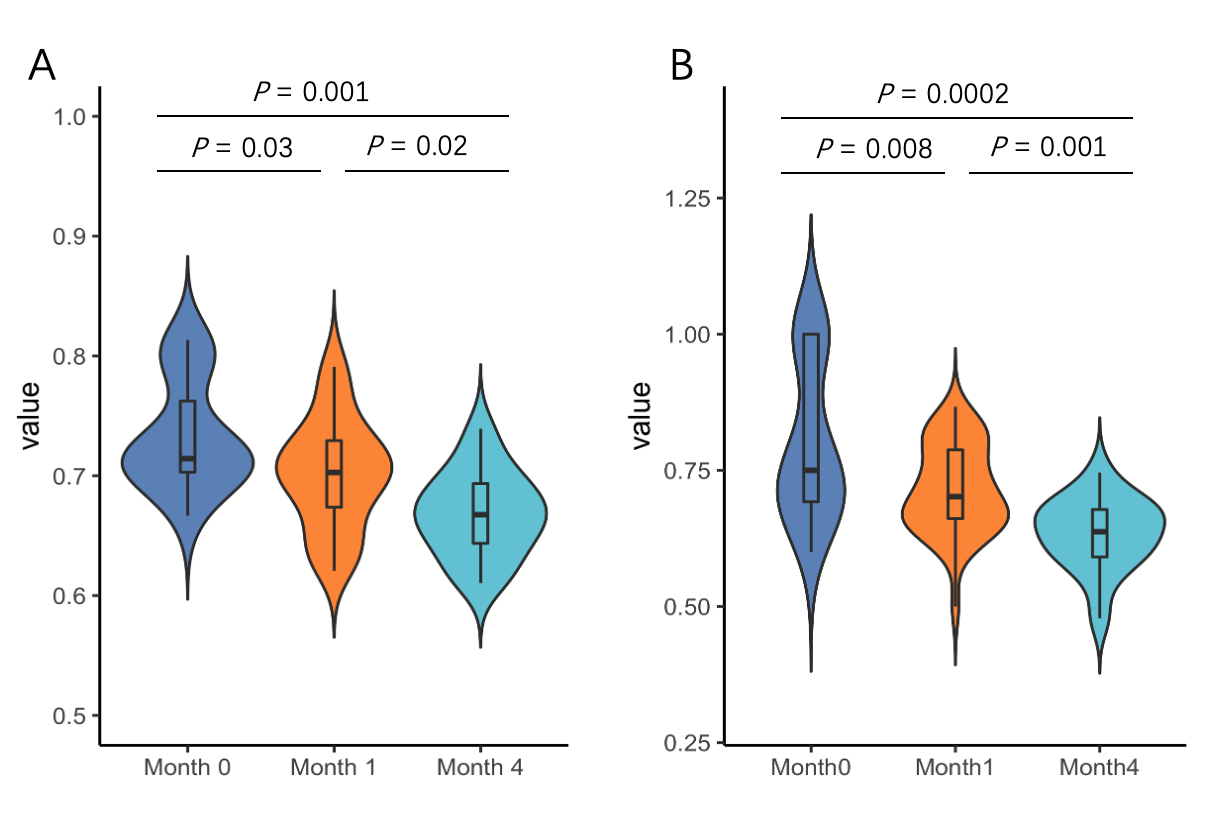}
    \caption{Violin plot at different months. (A) Predictions on all 2,291 phages. (B) Predictions on newly colonized phages. Y-axis: the percentage of temperate phages in each sample. }
    \label{fig:month}
\end{figure}

\paragraph{Ages} First, we group the samples by age and draw violin plot to show the lifestyle distribution in Fig. \ref{fig:month} A. Y-axis represents the percentage of temperate phages in each sample: $temperate/(temperate+virulent)$. We also record the p-value to show whether the two groups differ significantly. Fig. \ref{fig:month} A shows that as the baby grows, the percentage of temperate phages decreases, suggesting that more virulent phages colonize in the infants' gut with the infant's growth. Thus, we further investigate the lifestyle of newly colonized phages for each infant. Specifically, because we have three data samples representing \textit{month 0}, \textit{month 1}, and \textit{month 4}, we can first identify new phage contigs in month 1 by aligning reads of month 0 to the contigs in month 1 via Bowtie2 \cite{langmead2012fast}. If a contig in month 1 has no read mapping outputs, this contig is regarded as a newly colonized phage after month 0. Finally, the lifestyle prediction results on these newly colonized phages are shown in Fig. \ref{fig:month} B. The trend and conclusion are the same as Fig. \ref{fig:month} A and the p-value became smaller, indicating that the newly colonized viruses are more likely to be virulent phages.

Our results are consistent with the main conclusions of the original study \cite{liang2020stepwise}. During the early stage after birth, pioneer bacteria colonize the infant's gut. The prophages induced from these bacteria provide the predominant population. Then, as the infant grows, more phages and bacteria from the environment might reside within human guts too, leading to the change of the lifestyle composition.

\begin{figure}[h!]
    \centering
    \includegraphics[width=0.65\linewidth]{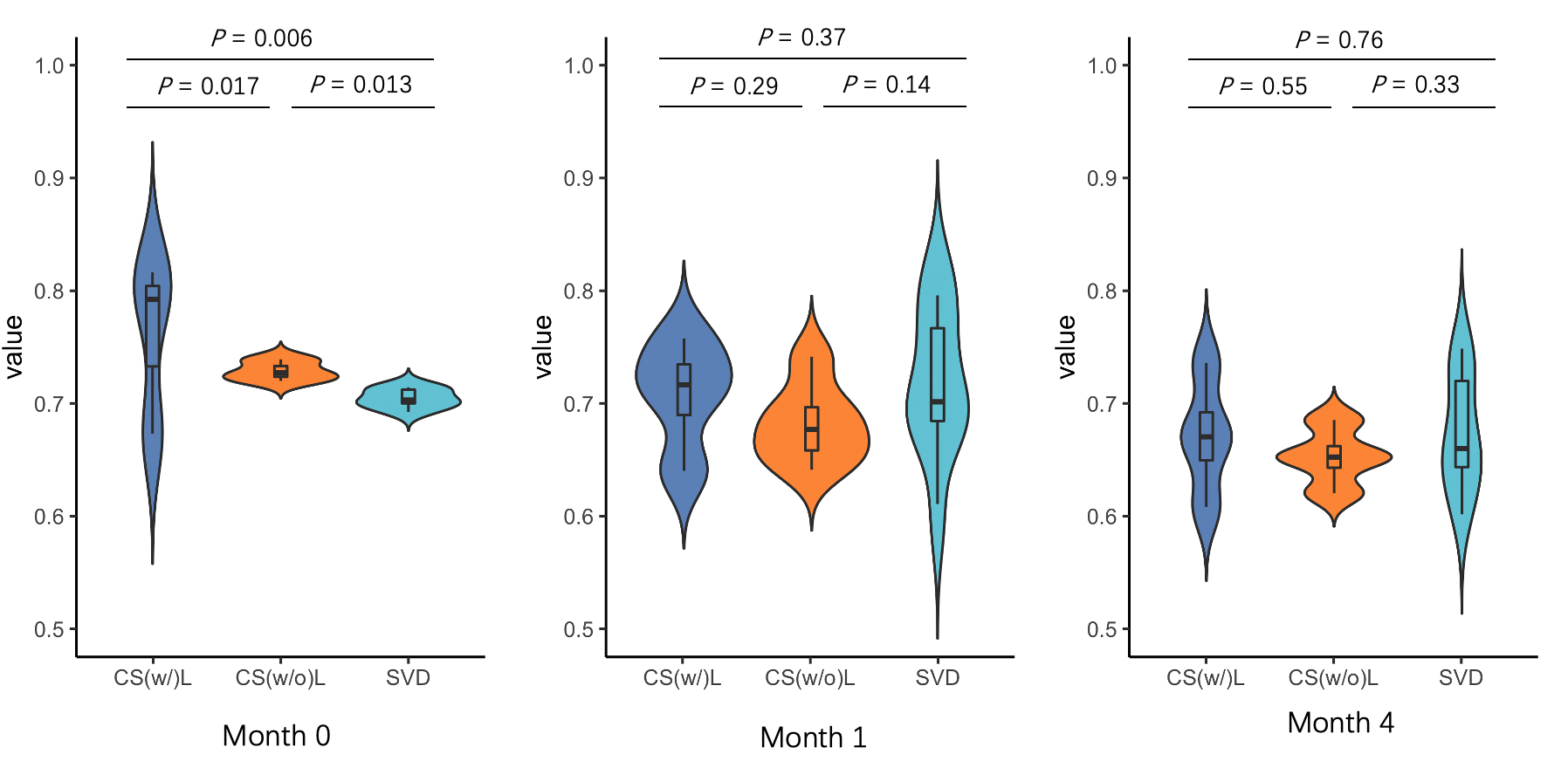}
    \caption{Violin plot in different delivery type: \textit{C-Section with labor (CS(w/)L)}, \textit{C-Section without labor (CS(w/o)L)}, and \textit{spontaneous vaginal delivery (SVD)}. To control variables, we group the samples according to the ages. The Y-axis represents the percentage of temperate phages in each sample. }
    \label{fig:delivery}
\end{figure}

\paragraph{Delivery type} Because the phage lifestyle on different age groups shows significant differences, we further investigate whether the phage lifestyle composition is influenced by the \textit{delivery} type. We first group the data by ages. Then, we draw the violin plots for three delivery types in each group. The results are shown in Fig. \ref{fig:delivery}.

There are three types of delivery: \textit{C-Section with labor (CS(w/)L)}, \textit{C-Section without labor (CS(w/o)L)}, and \textit{spontaneous vaginal delivery (SVD)}. In group \textit{month 0}, there is a significant difference between the delivery type and phage lifestyle composition. With the increase of the age, the difference gradually becomes smaller, suggesting that delivery type can influence the initial phage colonization of the infants. According to the explanation in \cite{liang2020stepwise}, the environmental contamination of different delivery types can affect the gut microbiome at \textit{month 0}, which might also lead to the difference.

\begin{figure}[h!]
    \centering
    \includegraphics[width=0.55\linewidth]{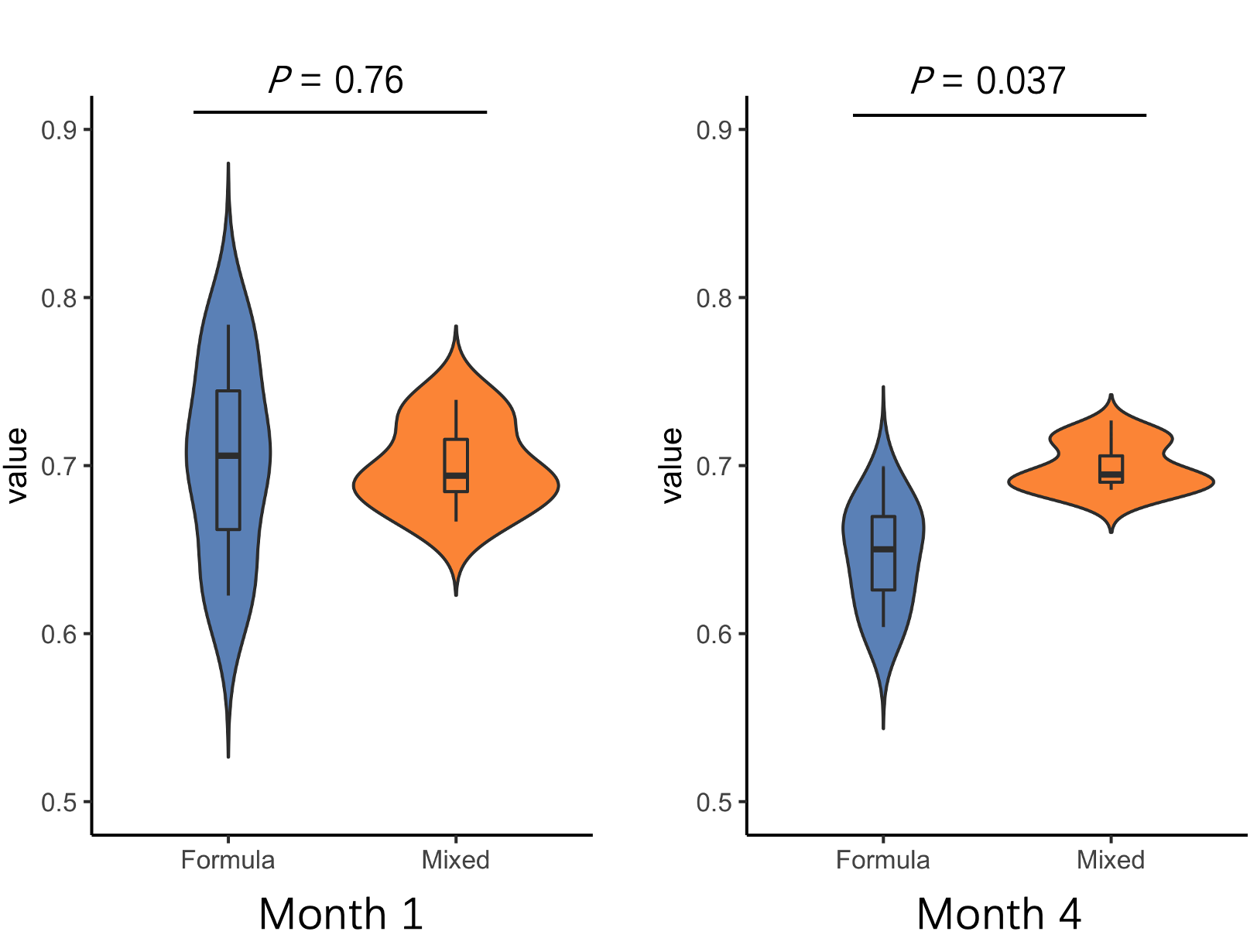}
    \caption{Violin plot in different feeding types. \textit{Formula}: the infants were fed with formula milk. \textit{Mixed}: infants were fed with both formula and breast milk. To control variables, we group the samples according to the ages. The Y-axis represents the percentage of temperate phages in each sample.}
    \label{fig:feeding}
\end{figure}

\paragraph{Feeding type} Because we only have feeding types for infants at \textit{month 1} and \textit{month 4}, we show the two groups' violin plots in Fig. \ref{fig:feeding}.  An interesting finding is that the feeding type does not show much difference at \textit{month 1}. But with the growth of infants, the proportion of temperate phages in the infants with formula milk decreases more rapidly than those with the mixed feeding type. One possible reason is that breastfeeding can reduce the chance of infection by microbes \cite{binns2016long, allen2005benefits, liang2021human} and thus, leading to relatively stable environments for the infants with access to breast milk.

\section{Discussion}
In this work, we propose a method named PhaTYP for phages' lifestyle prediction. As shown in the experiments, PhaTYP outperforms the available lifestyle prediction methods. We design two different tasks, a self-supervised learning task for learning protein association and a fine-tuning task for lifestyle prediction, to train the model so that PhaTYP achieves more stable performance on short contigs than other methods. In addition, we show an application of using PhaTYP to investigate the early viral colonization in human neonates. The prediction results of PhaTYP yielded several insights, most of which are consistent with recently published studies, suggesting that PhaTYP is a powerful means for studying and analyzing phage composition in metagenomic data.

Although PhaTYP has greatly improved phages' lifestyle prediction, we have several aims to optimize PhaTYP in our future work. One possible extension is to employ the knowledge distillation to reduce the parameter size of PhaTYP. Because of the stacking of transformer blocks in PhaTYP, it requires large computational resources (four 24Gb GPU units) to accelerate the training process. Distilling the knowledge in a neural network can provide a light-version model with the same performance for users who wish to re-train the model. Another extension is to incorporate phage identification \cite{shang2022accurate} in our software so that users can analyze phages' lifestyles in metagenomic data directly. 

In conclusion, based on our tests on both simulated and real sequencing data, PhaTYP achieves the most accurate lifestyle prediction among available tools. PhaTYP can be applied to various metagenomic data for analyzing virus compositions.

\section{Data Availability}
All data and codes used for this study are available online or upon request to the authors. The source code of PhaTYP is available via \url{https://github.com/KennthShang/PhaTYP}.

\section{Conflict of Interest}
There is no competing interest.

\section{Funding}
City University of Hong Kong (Project 9678241 and 7005453) and the Hong Kong Innovation and Technology Commission (InnoHK Project CIMDA).


\bibliographystyle{unsrt}  
\bibliography{references}

\end{document}